# The *Anna Karenina* principle:

# A concept for the explanation of success in science

First author and corresponding author:

Lutz Bornmann

Max Planck Society, Hofgartenstrasse 8, D-80539 Munich

E-mail: bornmann@gv.mpg.de

Second author:

Werner Marx

Max Planck Institute for Solid State Research, Heisenbergstrasse 1, D-70569 Stuttgart

E-mail: w.marx@fkf.mpg.de

**Abstract**

The first sentence of Leo Tolstoy's novel *Anna Karenina* is: "Happy families are all alike; every unhappy family is unhappy in its own way." Here Tolstoy means that for a family to be happy, several key aspects must be given (such as good health of all family members, acceptable financial security, and mutual affection). If there is a deficiency in any one or more of these key aspects, the family will be unhappy. In this paper we introduce the *Anna Karenina* principle as a concept that can explain success in science. Here we will refer to three central areas in modern science in which scarce resources will most usually lead to failure: (1) peer review of research grant proposals and manuscripts (money and journal space as scarce resources), (2) citation of publications (reception as a scarce resource), and (3) new scientific discoveries (recognition as a scarce resource). If resources are scarce (journal space, funds, reception, and recognition), there can be success only when several key prerequisites for the allocation of the resources are fulfilled. If any one of these prerequisites is not fulfilled, the grant proposal, manuscript submission, the published paper, or the discovery will not be successful.

**Key words**

Journal peer review, Bibliometrics, Scientific progress, *Anna Karenina* principle

# 1     Introduction

The first sentence of Tolstoy's novel *Anna Karenina* is: "Happy families are all alike; every unhappy family is unhappy in its own way" (Tolstoy, 1875-1877/2001, p. 1). Here Tolstoy means that for a family to be happy, several key aspects must be given (such as good health of all family members, acceptable financial security, and mutual affection). If there is a deficiency or failure in any one or more of these key aspects, the family will be unhappy. As there are a number of things required for a happy family and an unhappy family lacks a certain constellation of aspects (or at least one aspect), each unhappy family is unhappy in its own very specific way.

Diamond (1994; 1997) extended the principle behind Tolstoy's (1875-1877/2001) first sentence of the novel to understanding the requirements for success in complex undertakings, calling it the *Anna Karenina* principle (abbreviated here AKP). According to the AKP, for something to succeed several key aspects or conditions must be fulfilled. Failure in any one of these aspects leads to failure of the undertaking. That is, the success of complex undertakings always depends upon many factors, each of which is essential; if just one factor is lacking, the undertaking is doomed. The AKP is related to the conjunctive decision rule in consumer behavior research or psychology (Gilbride & Allenby, 2004), by which in a decision situation a person selects only those objects (the successful objects) that are found acceptable on all relevant criteria (a different decision rule is, for example, a compensatory rule: for an object to be successful, a negative, or unacceptable, rating on one criteria can be compensated by a positive, or acceptable, rating on another criteria).

Based on the AKP as stated by Diamond (1994; 1997), there are important implications for the conditions of success and failure: (1) Even though "we tend to seek easy, single-factor explanations for success" (Diamond, 1994, p. 4), there are usually not any for complex undertakings; (2) "Success actually requires avoiding many separate causes of failure" (Diamond, 1997, p. 157), and if only one cause of failure is avoided, there will be no success; (3) "No one property guarantees success, but many can lead to failure" (Shugan, 2007, p. 145); (4) As favorable outcomes require every detail to be right, whereas an unfavorable outcome only requires one wrong detail, "favorable outcomes are rare and more informative than unfavorable outcomes" (Shugan & Mitra, 2009, p. 11). However, this holds only for adverse environments: "In propitious environments, favorable outcomes convey less information (Shugan & Mitra, 2009, p. 13).

The AKP has been applied in many fields of research in recent years. Diamond (1994; 1997) uses it to find answers to the questions, "Why have so many seemingly suitable, big, wild mammal species, such as zebras and peccaries, never been domesticated, and why were the successful domesticates almost exclusively Eurasian?" (Diamond, 1994, p. 4). In accordance with the AKP, Diamond's theory names several necessary conditions related to geography for successful domestication (such as quick growth rate, no nasty disposition with tendency to kill humans). Moore (2001) applies the AKP to ecological risk assessments: "Following from the *Anna Karenina* principle, there are many ways to ruin an ecological risk assessment, but only a few pathways to success" (p. 236). McClay and Balciunas (2005) apply the principle to the area of biological control of weeds. In the field of empirical marketing research, Shugan (2007) formulated the "Anna Karenina bias" as follows:

> If we only observe survivors and survivors share the critical properties necessary for survival, then there will be little or no variation on the key variables (or constants) related to these properties. Hence, it will be difficult to infer the descriptive theory leading to success from the passive observation of survivors. We would need to actively observe nonsurvivors. (p. 146)

Shugan and Mitra (2009) applied the AKP in a paper on using averaging and nonaveraging statistics for success metrics: "When environments are adverse (e.g., failure-rich), non-averaging metrics correctly overweight favorable outcomes. We refer to this environmental effect as the Anna Karenina effect, which occurs when less favorable outcomes convey less information" (p. 4).

In this paper we use the AKP as a concept to explain the progress of success in science. Here we look at three central areas in modern science in which it is exceedingly difficult to be successful if resources are scarce: (1) peer review of research grant proposals and manuscripts (money and journal space as scarce resources), (2) citation of publications (reception as a scarce resource), and (3) new scientific discoveries (recognition as a scarce resource). If resources are scarce (journal space, funds, reception, and recognition), there can be success only when several key prerequisites for the allocation of the resources are fulfilled. If any one of these prerequisites is not fulfilled, the grant proposal, manuscript submission, the published paper, or the discovery is doomed to fail.

The AKP can yield important insights for understanding success in science, because it contradicts common assumptions. We frequently associate success with special characteristics of the successful persons and emphasize what is unusual about them (for example, see the interviews with highly-cited scientists at Thomson Reuters' ScienceWatch.com). When examining success, we mostly do not think about a lack of success. The AKP changes the focus. It throws light on the non-successful as an individual case that through its uniqueness is not successful. A successful thing is the standardized thing, which fulfills all requirements. In this way, for instance, the AKP could explain the tendency in peer review for making conservative decisions (Lamont, 2009). In the selection of grant recipients, a grant proposal is successful only if all of the funding organization's predetermined criteria are fulfilled; a risky, novel research approach outlined in a grant proposal is not successful in peer review in its own way, because it does not fulfill all criteria that the mainstream in a research area considers important.

It is not our intention in this work to show precisely which given set of prerequisites is necessary and sufficient to be successful in peer review of research grant proposals and manuscripts, citation of publications, and new scientific discoveries. This work introduces AKP to science studies only and describes some possible prerequisites for success in the three areas. It should be the task of the future empirical research to identify for each are the precise set of prerequisites.

## 2     Peer review and the *Anna Karenina* principle

Aside from the selection of manuscripts for publication in journals, the most common contemporary application of peer review in scientific research is for the selection of fellowship and grant applications. Peers or colleagues asked to evaluate applications or manuscripts in a peer review process take on responsibility for assuring high standards in various research disciplines. According to Lamont (2009), "peers monitor the flow of people and ideas through the various gates of the academic community" (p. 2). Although equals active in the same field may have limited vision due to their membership in the specialist group, they "are said to be in the best position to know whether quality standards have been met and a contribution to knowledge made" (Eisenhart, 2002, p. 241). Peer evaluation in research thus entails a process by which a selective jury of equals active in a given scientific field convenes to evaluate the undertaking of scientific activity or its outcomes. The jury of equals may be consulted as a group or individually, without the need for personal contacts among the evaluators. The peer review process lets the active producers of science, the experts, become the 'gatekeepers' of science (McClellan, 2003).

Other than the studies on reliability (i.e., is the selection of manuscripts or grant proposals reliable, or is it a chance result?), fairness (i.e., are certain groups of authors or applicants favored or at a disadvantage?), and predictive validity (i.e., does the process fulfill the objective to select the best manuscripts or proposals?) (Bornmann, 2011), there are few studies in the peer review research literature on the evaluation process as it occurs. Few studies deal with the typical review criteria of editors or program managers and with the editors' or program managers' rules for decision-making on these criteria that effectively yield the outcomes of the peer review process. Although many journals and funding agencies name a number of criteria that they use in their peer review process (see here the overview in Lamont, 2009), it is still unclear what criteria are in fact decisive for success and failure (see here Thorngate, Dawes, & Foddy, 2009). Sonnert (1995) indicated that "in stark contrast with the multi-faceted relevance of peer review in science, the peer review process has largely retained the characteristics of a 'black box'. It does produce quality judgments, but one does not quite know how they come about" (pp. 37-38). Gosden (2003) explained shortage of studies as follows: "As gatekeeping discourse, peer reviews remain largely under-researched principally due to their hidden status and issues of confidentiality" (p. 87).

In the following we would like to look at two studies published in recent years that dealt with the criteria or decision rules in peer review. The findings of both studies indicate that for success in the review process, the AKP plays a decisive role.

First, taking the journal *Angewandte Chemie International Edition* (AC-IE) as an example, Bornmann and Daniel (2009, 2010) investigated the editors' rules for decisions on acceptance or rejection of manuscripts as the outcome of the peer review process. Interestingly, the editors at AC-IE follow a "clear-cut" rule in most cases: the editors accept for publication only those manuscripts that the reviewers have evaluated positively regarding importance and print-worthiness in the AC-IE. Thus, a manuscript is published only if two reviewers choose the response category "very important" or "important" in answer to the question, "How important do you consider the results?" and also do not answer "no" to the question, "Do you recommend acceptance of the Communication?" The editors appear to deviate from this rule only rarely. If the editors decide on the basis of two external reviews, a manuscript must be reviewed positively by both reviewers to be accepted for publication. And if three reviews towards a decision are available to the editors, at least two must be positive.

Second, Bornmann and Daniel (2005) investigated the relationship of selection criteria and decisions in the committee peer review process used by an international foundation for the promotion of basic research in biomedicine, the Boehringer Ingelheim Fonds (B.I.F.), for awarding long-term fellowships to postgraduate researchers. For approval or rejection of fellowship applications, three criteria are decisive: According to Hermann Fröhlich, former managing director of the B.I.F., "in addition to the applicant's [track] record and the originality of the research project, there is a third element on which our judgment is based: the quality of the laboratory in which the applicant wants to pursue his project" (Fröhlich, 2001, p. 73). Bornmann and Daniel (2005) tested the extent to which approval of an application depends fundamentally on positive assessments on all of the three criteria using the Boolean probit statistical technique. This statistical procedure introduced by Braumoeller (2003) allows binary outcomes (here: 0 = rejection, 1 = approval) to be modeled as the results of Boolean interactions among independent causal processes. The results using this procedure confirm a conjunctural causation of the approval decision. In agreement with the prescriptive principles of the B.I.F. (Fröhlich, 2001), a positive assessment of the research project, in conjunction with both a positive assessment of the laboratory in question and a positive assessment of the applicant's achievement record, proves to have a highly statistically significant effect on approval of the application.

The findings of Bornmann and Daniel (2009, 2010) on AC-IE and Bornmann and Daniel (2005) on B.I.F. show clearly that for success in the peer review process, for the reviewers and decision-makers all criteria must be fulfilled. If reviewers at AC-IE assess a submitted manuscript as unimportant or if the B.I.F. board of trustees rates the laboratory where the grant applicant intends to conduct the research as not very favorable, they do not accept the manuscript for publication or approve the grant. In agreement with the AKP, for success in the peer review process all factors must be given; just one missing factor will lead to failure.

Through the increasing shift towards the 'soft-money system' in the financing of research and the use of papers as an evaluative measure of scientific performance (published by individual researchers or groups of scientists) in nearly all scientific disciplines, the peer review processes by research funding organizations and journals are confronted with an ever-increasing number of submissions (see, for example, Gölitz, 2008). This development poses new challenges to the processes. In earlier days, reviewers needed to reliably filter out proposals or manuscripts that did not meet a certain minimum standard (negative selection). In those days, the AKP most probably did not play a very great role in success in the peer review process: Since the resources were not too scarce, not all factors for the success of a grant proposal or submitted manuscript had to be given. Today, however, reviewers usually need to select the 'best' from a multitude of high quality scientific papers and proposals (positive selection). According to Yalow (1982), for today's peer review the question is "how to identify the few, those who make the breakthroughs which permit new horizons to open, from the many who attempt to build on the breakthroughs – often without imagination and innovation" (p. 401). And with these few, in accordance with the AKP, all factors must be given to be successful in today's highly competitive research environments.

## 3    Citations and the *Anna Karenina* principle

The aim of scientists publishing a paper is not only to present research findings to fellow researchers and to put their names on the findings (Merton, 1957) but also to invite fellow researchers "to freely take and use the information it [the paper] contains" (McClellan, 2003, p. 41). It complies with the rules of good scientific practice for scientists to accept the invitation and to relate their own perspectives, significant concepts, chosen methodologies, and definitions of problems to peers' perspectives, concepts, methodologies, and definitions. The method of indicating these relationships is through references or citations (Hooten, 1991). References link "documents and authors in accordance with the commonly perceived dynamics of knowledge production" (de Bellis, 2009, p. 14). In the area of research evaluation, if a paper has a high citation impact, very many citing papers have been "built upon" this cited paper. These highly-cited papers are called "hot papers," "fast breaking papers," or – if the citation impact sets in only after a certain period of time – "sleeping beauties."

However, the usefulness of citation counts for measuring research impact has been questioned (see here Joint Committee on Quantitative Assessment of Research, 2008). Already in the 1970s Garfield, founder of the Institute for Scientific Information (ISI) and its chairman emeritus, pointed out:

> Citation frequency is … a function of many variables besides scientific merit: an author's reputation, controversiality of subject matter, circulation, availability and extent of library holdings, reprint dissemination, coverage by secondary services, priority in allocation of research funds, and others. It is extremely difficult, even when possible, to clarify the relations among such variables and their relative impact on citation frequency. (Garfield, 1972, p. 476)

According to Gölitz (2005), other factors affecting citation counts are "the number of researchers currently working on that topic; articles on unfashionable or highly specialized topics, which can certainly be or become very important, are naturally less cited than articles on current, main-stream research" (p. 5539).

In recent years a number of studies investigated empirically the factors that have a significant effect on citation counts (Bornmann & Daniel, 2008). Montpetit, Blais, and Foucault (2008) collected data on the publication output of 758 Canadian political scientists and found:

> An article is more likely to be widely cited if it is published in a prestigious journal, if it is written by several authors, if it applies quantitative methods, if it compares countries, and if it deals with administration and public policy or elections and political parties. (p. 802)

Lansingh and Carter (2009) identified the following variables as having the greatest effect on citation counts: number of authors, country/region of publication, subject area, language, and funding. A very large-scale study (Vieira & Gomes, 2010) looked at the whole set of the more than 220,000 articles published in the year 2004 referenced in Thomson Reuters' Web of Science for Biology & Biochemistry, Chemistry, Mathematics and Physics. The researchers concluded: "The number of co-authors, the number of institutional addresses, the number of pages, the number of references and the journal impact factor were considered as basic features that may have direct influence on the citation count" (Vieira & Gomes, 2010, p. 11).

The results of these and several other studies (see an overview in Bornmann, Mutz, Neuhaus, & Daniel, 2008) indicate that the research activities of scientists, publication of their findings, and citation of the publications by colleagues in the field are mainly also social activities. This means that citation counts for the scientists' publications are not only an indicator of the impact of their scientific work on the advancement of scientific knowledge but also the result of many factors besides the scientific quality of the scientists' research. Against this backdrop, two competing theories of citing have been developed in past decades, both of them situated within broader social theories of science. One is often called the normative theory of citing and the other the social constructivist view of citing. Both of these citation theories attempt to clarify the fundamental question as to why author $x$ cited article $a$ at time $t$ (Sandström & Sandström, 2008).

The normative theory, following Merton's (1973) sociological theory of science, basically states that scientists give credit to colleagues whose work they use by citing that work. Thus, citations represent intellectual or cognitive influences on scientific work. Merton (1988) expressed this aspect as follows:

The reference serves both instrumental and symbolic functions in the transmission and enlargement of knowledge. Instrumentally, it tells us of work we may not have known before, some of which may hold further interest for us; symbolically, it registers in the enduring archives the intellectual property of the acknowledged source by providing a pellet of peer recognition of the knowledge claim, accepted or expressly rejected, that was made in that source" (p. 622) (see also Merton, 1957; Merton, 1968).

The social constructivist view of citing is grounded in the constructivist sociology of science (see, e.g., Collins, 2004; Knorr-Cetina, 1981; Latour & Woolgar, 1979). This view casts doubt on the assumptions of normative theory and questions the validity of evaluative citation analysis. Constructivists argue that the cognitive content of articles has little influence on how the articles are received. Scientific knowledge is socially constructed through the manipulation of political and financial resources and the use of rhetorical devices (Knorr-Cetina, 1991). For this reason, citations cannot be satisfactorily described unidimensionally through the intellectual content of the article itself. The probability of being cited depends on factors that do not have to do with the accepted conventions of scholarly publishing. Gilbert (1977), who has been particularly associated with the social constructivist view, saw citations as an aid to persuasion, finding that scientists prefer to cite documents that are supportive of what they write and preferably written by noted experts.

Whereas Cronin (1984) found the existence of two competing theories of citing behavior hardly surprising, as the construction of scientific theory is generally characterized by ambivalence, for Liu (1997) and Weingart (2005) the long-term oversimplification of thinking in terms of two theories reflected the absence of one satisfactory and accepted theory on which better informed use of citation indicators could be based. Liu (1997) and Nicolaisen (2003) saw the dynamic linkage of the two theories as a necessary step in the quest for a satisfactory citation theory. Four studies investigated the validity of the two theoretical approaches empirically. In agreement with the constructive view of citing behavior, Collins (1999) suggested that political and economic forces within the research process led to some papers being ignored by scientists, whereas some were picked out. In contrast, studies by Baldi (1998), Stewart (1983), and White (2004) provided more support for a normative interpretation of the allocation of citations than for a social constructivist interpretation. Implications of the study by Baldi (1998) have been discussed extensively in the literature in recent years. Cronin (2004) assessed the Baldi study as "an important and methodologically rigorous study" (p. 44). Borgman and Durner (2002) found further comparison of citing behavior within different disciplines necessary if we are to determine how far Baldi's results may be generalized. Likewise, Small (1998) was not completely convinced by the results and stated: "a direct empirical test of the two theories seems difficult, and we need to take a step back and view these two theories in a broader context" (p. 143).

That broader context could be the AKP. Although most empirical studies that examined factors influencing citations (see above) showed that there are always several factors having an important impact (such as number of authors of a publication, quality of a work, and reputation of the journal in which the paper is published), the two citation theories both refer to a limited range of factors. The normative theory stresses factors that relate to the quality of a work, such as the intellectual content of a paper, and the social constructivist view of citing emphasizes factors that have nothing to do with academic publishing conventions (such as citations as a way to make the content convincing to the reader). Depending on the theory used, a successful paper that has achieved a high citation impact is attributed with specific attributes that have led to its success.

With the AKP the two citation theories can be brought together in a broader context: According to the AKP, a paper is successful and will be highly cited if a number of key factors are fulfilled, one (important) factor being the quality of the paper. Another factor is, for instance, interest in a paper will be increased if at least one of the authors is well-known. Also, a short and succinct paper with an appealing title is good for reception among colleagues. The extent to which these and other criteria (both quality criteria and reputation/circulation criteria) are fulfilled should be reflected in the citation counts of scientific publications. If one of these factors is lacking (for example, a high-quality paper is not published in a prestigious international journal), the citation frequency of the paper will most probably not be above average.

In the area of citation counts, the AKP is related to what is called the principle of anti-diagnostics, a medical term that Braun and Schubert (1997) applied to the area of bibliometrics:

> While in medical diagnosis numerical laboratory results can indicate only pathological status but not health, in scientometrics, numerical indicators can reliably suggest only eminence but never worthlessness. The level of citedness, for instance, may be affected by numerous factors other than inherent scientific merits, but without such merits no statistically significant eminence in citedness can be achieved. (p. 177)

The meaning of this principle is that citation counts tend to indicate high quality of scientific papers better than they indicate low quality. A paper that has not received many citations is not automatically a paper of low quality. The reasons for the lack of citations can be many (see above).

## 4 The *Anna Karenina* principle in scientific progress

It is still not clear today how scientific progress functions in detail and what the prerequisites for scientific breakthroughs are. According to philosopher of science Thomas Kuhn, the development of science takes place in a cyclical pattern: Kuhn (1962) made a distinction between "normal science," which is oriented towards recognized explanatory models and methods (paradigms), and scientific revolutions, which become necessary when puzzling deviations can no longer be solved by existing paradigms. These anomalies first lead to model drift and then to model crisis, when "extraordinary science" begins. Here the existing paradigms become blurred and are then replaced by new paradigms (paradigm change). These are the milestones of research, or scientific revolutions. They start up a new phase of normal science, but without cumulatively approximating closer and closer any absolute truth. To illustrate, Kuhn (1962) used the metaphor of biological evolution, which is similarly driven by problem solutions and does not advance towards any fixed goal. According to Kuhn (1962), the old and new paradigms are incommensurable – that is, there is no mutual understanding between the old and new ways of looking at things and between the old and new terminology. Although several natural scientists have criticized the incommensurability thesis (e.g., Weinberg, 1998), the importance of Kuhn's (1962) theory has never been fundamentally called into question.

Whereas Kuhn (1962) described that there is a turn of the tide from normal science to scientific revolution, he did not describe what the prerequisites must be for a revolution to take place:

> Rather we must explain why science – our surest example of sound knowledge – progresses as it does, and we must first find out how, in fact it does progress. Surprisingly little is yet known about the answer to that descriptive question. A vast amount of thoughtful empirical investigation is still required. (Kuhn, 1977, p. 289)

But by naming important criteria for assessing the quality of new theories, Kuhn (1977) already discussed what is basically needed for an old theory to be replaced by a new theory:

> What, I ask to begin with, are the characteristics of a good scientific theory? Among a number of quite usual answers I select five, not because they are exhaustive, but because they are individually important and collectively sufficiently varied to indicate what is at stake ... These five characteristics – accuracy, consistency, scope, simplicity, and fruitfulness – are all standard criteria for evaluating the adequacy of a theory. If they had not been, I would have devoted far more space to them in my book, for I agree entirely with the traditional view that they play a vital role when scientists must choose between an established theory and upstart competitor. Together most others of much the same sort, they provide *the* shared basis for theory choice. (pp. 321-322)

However, these basic criteria for assessing theories are too abstract to allow us to name the prerequisites for paradigm change in a discipline and to more closely examine the process of scientific progress, especially under the complex conditions of modern research.

Kuhn (1962) fully explained the complexity of scientific revolutions taking the example of the Copernican Revolution. Kragh and Smith (2003) underlined the complexity also in connection with modern cosmology – the discovery of the expanding universe:

> Is it at all reasonable to ask who made this discovery, and when? The expansion of the universe became recognized as a fact, but can it be described as a discovery? If it can, who deserves the credit, and for what? … These questions are far from simple. It is commonly accepted among historians and philosophers of science that most discoveries cannot be neatly localized in space and time. They are not individual events, but complex and often messy processes extended over a period of time and involving many actors. Many scientific discoveries consist of several, more or less connected in sights that in the end result in a consensus as to how the discovery has been made. A discovery does not necessarily require a discoverer or a discovery event. (p. 142)

Here Kragh and Smith (2003) pointed to something that we would like to clearly define in this paper. We will attempt in the following to name some prerequisites for scientific revolutions, adhering closely to the complex process of the development of science. It is not our intention to formulate the precise set of prerequisites, but to introduce the AKP here (see the Introduction). We would like to bring in the AKP as the overarching concept, because via the step by step fulfillment of several prerequisites it delivers a kind of explanation of a scientific breakthrough. In other words, the principle does not contain any previously overlooked new conditions for scientific revolutions but instead delivers a certain view of, or organizing schema for, that which leads to a breakthrough. The starting point of scientific activity that develops into a scientific revolution is usually efforts to understand fundamental connections and to answer the associated core questions (for instance, how did the surface of the earth or the universe come into being?). Chance discoveries often play an important role in these activities, in that they trigger events or drive development forward (Merton & Barber, 2004). But the prerequisites for scientific revolutions are always factors like independently produced empirical data that lead to similar results, the formulation of sound hypotheses that can be tested, and the development of a plausible theory that others acknowledge to provide an explanation of a phenomenon. Only when these and further prerequisites are fulfilled is there a "happy family" (in this case, a scientific revolution), which is like all other happy families in that all prerequisites are fulfilled.

We would like to regard the AKP as supplementing or completing Kuhn's (1962) theory – the change from phases of normal science to scientific revolutions. The principle does justice to the complexity of scientific progress, in that it resolves the process that leads to a scientific revolution into individual steps or prerequisites. Additions to Kuhn's (1962) theory seem necessary also because Kuhn used examples from earlier centuries (with the exception of the rise of quantum physics at the beginning of the twentieth century). Kuhn's theory with its two phases is accordingly kept relatively simple. Modern science, especially in the age of Big Science starting around 1960, is characterized by substantial changes in "science culture," as seen among other things in the number of researchers and publications, the type and extent of dissent and cooperation among scientists, and the establishment of large-scale research projects and institutions. For this reason we think it is necessary to extend Kuhn's (1962) theory in view of modern research. For an improved understanding of revolutions in modern research, an extended theory is needed.

In the following, with a view to basing the discussion of AKP on typical examples taken from modern science, we refer repeatedly to some of the most significant scientific revolutions in the twentieth century in the area of the core disciplines in the natural sciences: the development of modern cosmology (example 1), the development of modern geophysics in the form of plate tectonics (example 2), and the beginnings of quantum physics (example 3). The starting points in these areas of research were some of the core questions in the natural sciences: How did the cosmos develop? What is matter made of? How do the chemical and physical interactions of matter work? What is the chemical basis of metabolism and the heredity of living things? What forces and events shaped the surface of the earth?

Before we come to some prerequisites that should be fulfilled for scientific revolutions, the three examples will be introduced in brief.

Example 1: Modern cosmology began at the start of the twentieth century with Albert Einstein's (1905b, 1915, 1916, 1917) special and general theories of relativity, which revolutionized our understanding of space and time and introduced a new theory of gravitation. In the 1920s mathematical considerations led to the finding that the paradigm of a static universe can be replaced by the paradigm of a dynamic universe (Friedmann, 1922, 1924; Lemaitre, 1927). Already before this, initial evidence had been found that called the notion of an eternally static universe into question: Slipher (1912, 1917) found the redshift of the stellar spectral lines, which was the starting point for the discovery of the receding galaxies by Hubble (1929) and Hubble and Humason (1931). Hubble and Humason combined a new method of cosmological distance measurement with Slipher's data and also with their own measurement of the redshift and found a linear relationship between distance and velocity of the galaxies.

The cosmic microwave background radiation discovered by Penzias and Wilson (1965) was immediately interpreted as a relic of an event when matter as well as time and space came into being. In connection with the discovery of the redshift, this finally led to the paradigm of the Big Bang and an expanding universe. The origin and the frequency distribution of the chemical elements and the distribution of radio galaxies fit the Big Bang model very well. However, it took a long time to verify an important prediction of the model: The uneven distribution of matter in the universe in the form of a strong concentration of matter in isolated galaxies demanded a slight irregularity already at the beginning of the evolution of the cosmos. These beginning fluctuations in the density of matter must have been imprinted on the cosmic microwave background radiation and thus be provable as a pattern of small temperature differences. When measurements by satellite revealed just this pattern in the 1990s, the Big Bang model was finally confirmed and became today's recognized cosmological standard model (see Marx & Bornmann, 2010).

Example 2: Modern geophysics began at the beginning of the twentieth century with Alfred Wegener. Examining the question of the origin of the continents, Wegener found that there is a close fit between large-scale geological features of the coastlines of separated continents. The matching features were an indication that the continents may have been joined together at one time. Because Wegener (1912, 1915) could not present any plausible mechanism for his theory of "continental drift," it at first found little support among professional colleagues. Further evidence for continental drift was found starting only in the mid 1950s, with paleomagnetic work on the direction of magnetization of rocks (Mason, 1958; Mason & Raff, 1961; Runcorn, 1955). This direction was apparently caused by the movement (spreading) of the sea floor in connection with periodic polar reversals of the earth's magnetic field, which gave the rock its magnetic direction. In the early 1960s Hess (1962), and independently also Dietz (1961), developed a new theory of mantle dynamics, which was based on thermal convection currents in viscous mantle rock.

The rock magnetism data and the proposed mechanism were brought together in the Vine-Matthews hypothesis on the mobility of the earth's crust (Vine & Matthews, 1963). Wilson (1963, 1965) produced further indications, independent of the magnetic data, and expanded the theory of the moving surface of the earth considerably. Wilson was the first to name the masses of moving with convergent boundaries "plates," and he divided the earth's outer layer into at least six large and several smaller moving plates, which are kept in motion by convection currents in mantle rock. Seismology (the science of earthquakes) helped the plate tectonics model achieve the final breakthrough, in that it showed that the boundaries of the postulated plates are identical to known earthquake zones. For the first time there was now a complete picture regarding the events and forces that shaped the surface of the earth. With plate tectonics, a new field of research arose that dealt with large-scale motions of the earth's crust.

Example 3: Quantum physics began with the discovery of the quantum of action (*Wirkungsquantum*) by Max Planck. The quantum of action is an important physical constant; it says that physical quantities, such as energy, can take on only discrete values. Max Planck (1900a, 1900b) discovered it but restricted quantization to the exchange of energy between matter and electromagnetic field. Planck saw quantization as a temporarily necessary mathematical trick to allow theoretical treatment of the spectrum of black-body radiation (the color of hot objects such as electric light bulbs). Einstein (1905a) applied the concept of quantization to light itself and, with his light quantum hypothesis, he convincingly explained the photoelectric effect, which is the emission of electrons from matter as a consequence of their absorption of energy from light. For that explanation, Einstein received the Nobel Prize in Physics in 1921. Based on Einstein's hypothesis of light quanta, quantum theory became the basis of quantum mechanics in the mid 1920s. Quantum mechanics describes the behavior of energy and matter at the atomic and subatomic scales (that is, atomic and nuclear physics) and became one of the main pillars of modern physics. Quantum electrodynamics (QED) developed out of quantum mechanics in the 1940s. Today, QED is the accepted paradigm of electromagnetic radiation and its interaction with charged particles.

In the following, with the aid of these three examples, we want to present some of the conditions that we assume must be fulfilled for a scientific revolution to occur in a field. As prerequisites, solid evidence in answer to basic questions must be presented (1), which is taken up by colleagues (2) and can be verified by means of independent data and methods (3). Especially important is the statement of verifiable predictions that can then be confirmed (4). This requires techniques for gathering data (5). Of decisive importance are satisfactory and convincingly formulated paradigms that answer more questions than they raise (6). In demand in this connection are simplicity, elegance (aesthetic qualities) (7) and explanatory power of the new paradigm (8). For understanding of the paradigm, clear and plain language and the introduction of easy to remember labels are very helpful (9). Finally, what is needed is the last crucial step that leads to the definitive breakthrough and the establishment of a new paradigm (10). For a scientist making a crucial contribution to a scientific revolution, we find important stubbornness in thinking as well as good networking among colleagues in the field (11). The importance of these prerequisites for the occurrence of scientific revolutions will be illustrated in the following using the three examples outlined above.

Prerequisite 1: Solid evidence

Efforts to answer core questions in science are not only driven forward by not understood facts and connections but also steered in a new direction via solid evidence (empirical evidence, or initial evidence not yet confirmed independently). Submitting this evidence and interpreting it in view of a possible new paradigm to explain a phenomenon open up the matter to scientific dispute. The initial evidence that could speak for a new paradigm is usually still quite vague and not abundant. The postulated receding of the galaxies (example 1) was at first based exclusively on observation of the redshift in the emission spectra of stars in the galaxies; the only empirical proof of the continental drift hypothesis (example 2) was for a long time the measurement of rock magnetization. The technical options for gathering further evidence are often still underdeveloped, which is why independent confirmation at first fails to appear. Initial evidence that speaks for a new paradigm is frequently questioned, because for the most part there is a lot of room for interpreting it within the framework of the old paradigm. For instance, the discovery of the redshift (example 1) could also be explained within the paradigm of a static universe, as de Sitter (1917) proposed. This De Sitter's model was based on a different geometry than that of Einstein's (1915, 1916, 1917) also static universe, and it seemed to allow for redshift without requiring a new paradigm of an expanding universe. Analogously, the direction of magnetization of rocks could have been changed by chemical processes (example 2), so that it had little to do with the magnetic field of the earth at the time the rock was formed (and thus little to do with rock moving).

<u>Prerequisite 2: Interest among colleagues who take up on the ideas</u>

When Hubble (1929) and Hubble and Humason (1931) published the linear relation between distance and velocity of galaxies, discussion was opened among colleagues in the field (and beyond) on the paradigm of a dynamic universe (example 1). This is shown among other things by the relatively great number of citations of these papers (Marx & Bornmann, 2010). Following publication of Wegener's (1915) book on continental drift, reactions came quickly, even if they were at first overwhelmingly critical (example 2). The pioneering works of Planck (1900a, 1900b) and Einstein (1905a) were the start of a discussion still ongoing today on the physical laws in the microcosm and their philosophical aspects (example 3). Outstanding discoveries are usually soon taken up by colleagues, and they often create excitement, creating an atmosphere of departure (or a scientific "gold rush").

But that is not always the case. Sometimes new evidence or ideas are at first noted and not followed or pursued further. Whether or not new ideas are taken up depends essentially on whether they are put forward in an already existing discussion (discussion on the core questions in a field, for example) or whether they are difficult for colleagues in the field to interpret. In a discipline, reserved judgment on ideas that can lead to a new paradigm is not least due to the fact that science must necessarily be conservative so as not to have to turn constantly to new paradigms. It is thus not always blindness or obstinacy that sometimes makes researchers react cautiously (or not at all) but rather practicality.

For instance, Einstein's (1905b) new interpretation of space and time and his new theory of gravitation (1915, 1916, 1917) at first overtaxed many colleagues in the field (example 1). The discovery of the quantum of action by Planck (1900a) was also at first very disconcerting to colleagues (and to Planck himself as well) (example 3). The discovery of the redshift by Slipher (1912, 1917) (example 1) and the magnetic lineations or stripes on the ocean floor by Mason (1958) and Mason and Raff (1961) (example 2) are further good examples of overtaxing: At the time of the discoveries, these researchers and their colleagues did not know quite what to make of them. But the discoveries found their way into the scientific archives (the professional journals and databases) and could be used later for the research.

<u>Prerequisite 3: Verifying evidence from independent research groups</u>

Indications and initial evidence are usually based on still sparse data. Further evidence found by independent research groups is an important prerequisite for a field to develop a new paradigm. In Kuhn's (1977) later work, he comments on how this evidence helps in the choice between competing theories:

> Measurement can be an immensely powerful weapon in the battle between two theories, and that, I think, is its second particularly significant function. Furthermore, it is for this function – aid in the choice between theories – and for it alone, that we must reserve the word "confirmation." (p. 211)

For Kuhn (1977) new empirical data and continuous theory development are essential for progress in a field:

> One recurrent implication of the preceding discussion is that much quantitative research, both empirical and theoretical, is normally prerequisite to fruitful quantification of a given research field. In the absence of such prior work, the methodological directive, "Go ye forth and measure," may well prove only an invitation to waste time. (p. 213)

Not uncommonly, new evidence pertinent to answering fundamental questions has been produced by different researchers independently, at the same time (or also at different times) and in separate lines of work (sometimes even in different, unrelated subdisciplines). For instance, at the start of the development of plate tectonics (example 2) both land and ocean based magnetic measurements were made. At a later point in time, the different lines of research were combined. This happens frequently if a researcher having a broad, interdisciplinary overview recognizes the lines of work and uses them as the basis for a new paradigm. Of course, the prerequisite for this is that a number of pieces of evidence, produced in different subdisicplines based on different experimental methods, point in the same direction.

With data from independent research groups, there is a much higher probability that a new paradigm based on initial evidence is in fact valid. However, the independent evidence must be recognized as the elements that can confirm a theory. This is not always the case immediately. In geophysics (example 2) the sea floor magnetic stripes had not at all been looked for in a targeted manner after the discovery of the remanent magnetization of rocks on land (Creer, Irving, & Runcorn, 1957; Runcorn, 1955) but were instead found completely independently of that discovery and rather by chance (Mason, 1958; Mason & Raff, 1961). The connection between these observations and suppositions concerning crustal mobility were recognized only later. However, sometimes in efforts to verify a (new) paradigm, there is a deliberate search for evidence that can confirm or contradict the paradigm. For instance, the spatial irregularity of cosmic microwave background radiation (the insignificant fluctuations in temperature that the theory of a dynamic universe requires) (example 1) was not found by chance but instead systematically in the framework of an expensive satellite program.

<u>Prerequisite 4: The paradigm should make possible (correct) predictions</u>

The usefulness of a paradigm is not least measured in terms of the extent to which it can yield correct predictions. There is a good example of this in cosmology (example 1): Cosmic microwave background radiation was found by chance by Penzias and Wilson (1965). However, it had been predicted some years previously, and Alpher (1948) and Alpher and Herman (1949) had even calculated its wavelength. In this connection the prediction of fluctuations in the background radiation is no less impressive than the successful empirical observation. Geophysics (example 2) also provides an example here: The Canadian geophysicist Wilson (1963) supposed that the age of the islands of Hawaii increases with their distance from the East Pacific Rise. An essentially stationary hotspot of erupting magma had apparently created a trail of volcanic islands, which then drifted with the moving crust, and their active volcanism gradually stopped. Determination of the age of rock on the islands by radiometric dating confirmed Wilson's assumptions completely. An expedition of the drilling ship Glomar Challenger measured the age profile of the sea floor over the Mid-Atlantic Ridge systematically and comprehensively. The result: The age of the sea floor, as expected, increases symmetrically with increasing distance from the ridge to the west and east. This was the clinching evidence that new magma from deep within the earth rises at the structurally weak joins between the plates that make up the earth's crust and eventually erupts along the crest of the ridges to create new, spreading oceanic crust, transporting the continents along with it.

<u>Prerequisite 5: Suitable techniques for the required measurements</u>

Sometimes there are no techniques or technologies available for exact empirical testing of hypotheses that could lead to a new paradigm. Measurement of irregular cosmic microwave background radiation, which would concur with an expanding universe, became possible only with space travel (example 1). The demonstration of inhomogeneity in the microwave background radiation was then a convincing confirmation of the Big Bang model.

Without the existence of these fluctuations, it would not be possible to explain the unequal distribution of matter in the universe, the birth of galaxies, and thus our own existence. The developments in plate tectonics were very closely connected with the technical advances of wartime (Second World War) and the postwar period (example 2). Probes such as the echo sounder and magnetometer were decisive prerequisites for large-scale measurements of the structure and magnetization of the sea floor. Besides the mechanism responsible for continental drift, the result of the drift had to be investigated: The predicted slow drift of the continents could only be measured precisely, and then finally proven, only later through the use of satellites.

<u>Prerequisite 6: A theory that provides a plausible explanation of the empirical findings</u>

In the interplay of theory and experiment that essentially shapes scientific advancement, what is always needed in addition to evidence in the form of empirical data is a plausible interpretation of that data in a (new) theory. The theory must not only describe but also explain how something works (such as, how the earth's continents move). Thus the theory must contain the exact mechanism in the form of driving powers and processes. Regarding the hypothesis of continental drift (example 2), Wegener's approach (1912, 1915) was not recognized by colleagues in the field, because Wegener had no convincing mechanism for how continents move; he thought that tidal forces caused by the moon might be responsible. However, that was not possible quantitatively. A convincing reason for continental drift was presented only a half century later by Hess (1962): material heated by radioactive elements in the earth's interior and the resulting convection currents. As Hess's proposal that the sea floor itself moved, or spread, was unorthodox at the time, he cautiously called his theory "geopoetry."

With the overwhelming evidence and many premises it was only a question of time before an unbiased (young) researcher (Fred Vine) brought together Hess's (1962) theory and the already available evidence (Mason, 1958; Mason & Raff, 1961) in a convincing paradigm:

the Vine-Matthews hypothesis (Vine & Matthews, 1963). The hypothesis brought together the idea of sea floor spreading and the observed phenomenon of the magnetic striping patterns. This interplay of theory and empirical data connected up previously separate lines of work, thus preparing the ground for the later development of plate tectonics. Many other researchers also working in these areas had not been as versed in both geology and physics as were Vine and Matthews and had been viewing the connections from too narrow a perspective. The Vine-Matthews hypothesis was an important prerequisite for further progress in geophysics, as it was a plausible and solid theory of the supposed, very slow movement of the earth's crust.

Prerequisite 7: The paradigm is simple and elegant

The persuasive power of a new paradigm has a lot to do with whether it is simple and elegant (Kuipers, 2002; McAllister, 1996). Invited to write a self-chosen inscription on a blackboard after holding a lecture at the University of Moscow in 1956, Paul Dirac wrote, "A physical law must possess mathematical beauty" (Kragh, 1990, p. 275). Dirac (1928a, 1928b) used this premise when he derived the existence of the antielectron from his wave equation for the electron (the Dirac equation) and his hole theory based on it. The antielectron was discovered shortly thereafter by Anderson (1932) in the cosmic radiation (example 3). The Big Bang model (example 1) provided an elegant explanation of the origin and structure of the visible universe. The paradigm of plate tectonics (example 2) is very satisfactory aesthetically, as it provides simple and compelling answers to most of the basic questions in the geosciences.

Prerequisite 8: The paradigm has great explanatory power

Prior to the 1960s there was no generally accepted paradigm for the evolution of the surface of the earth (example 2). There was no satisfactory answer to the question of the origin of continents, oceans, mountains, valleys, and volcanoes. The answers came only starting in the 1960s with the paradigm of plate tectonics, which found general acceptance around 1970. The paradigm came about as the result of bringing together various subdisciplines of geophysics. The many branches of science that make up study of the earth (geomagnetism, seismology, petrography, and geophysics) remained at first fragments. It was the new paradigm of plate tectonics that brought them together (Anderson, 1971). The explanatory power of the Big Bang model (example 1) results from the many points of contact with classical astronomy and high energy physics. The scientific community recognized the great explanatory power of the Big Bang very early on.

Prerequisite 9: The paradigm has a catchy name

Appealing names for a new paradigm (such as Big Bang or plate tectonics) make it easier for the scientist (and persons outside of science) to name things and put things in a nutshell. Authors that introduce attractive names in the specialist literature do the discipline a good turn. A catchy name is important, because when a new paradigm is being established there is not yet any generally accepted, uniform language usage. In the debate over the Big Bang model versus the competing Steady State model (example 1), the term Big Bang was coined, rather disparagingly, by an opponent of the theory (Fred Hoyle) during a talk on a BBC radio program in 1950. The catchy phrase caught on, in both camps, and it got right to the heart of the new paradigm. When Wilson (1965) published a paper on the transform faults (plate boundaries, at which the plates move in relation to one another) (example 2), he named the moving rock masses "plates." Wilson divided the earth's surface into several major and minor plates that are kept in movement by convection currents in the earth's crust. In the paper Wilson (1965) gave the new paradigm an attractive name, calling it "plate tectonics."

Prerequisite 10: The last crucial step is achieved

In connection with the history of modern cosmology (example 1), the question arises as to why Hubble (1929) and Hubble and Humason (1931) are generally acknowledged to be the discoverers of the receding of the galaxies and not, for instance, Slipher (1912, 1917), who much earlier had found decisive initial evidence. Although they did not venture to interpret their findings as demonstrating an expanding universe, Hubble (1929) and Hubble and Humason (1931) had achieved the crucial last step and presented their findings clearly and unmistakably in words and images (in contrast to their forerunners):

> But surely it is true that Hubble's measurements, added to the data of Slipher (and Humason) provided the crucial last step to finally put together the first full recession graph? In other words, Hubble's work may have formed only part of the contribution, but it was the crucial last brick – a common occurrence in science! (C. O'Raifeartaigh, personal communication, October 17, 2010).

In analogous fashion, publication of the Vine-Matthews hypothesis (Vine & Matthews, 1963) can be seen as the crucial last step for establishing the paradigm of plate tectonics (example 2). The papers by Hubble (1929) and Hubble and Humason (1931), and Vine and Matthews (1963) have a strong synthesis character, as they bring together previously separate lines of research: In Hubble's case the measurement of cosmic distance and measurement of the spectral shift of galaxies, and in Vine and Matthews the striped pattern of natural remanent magnetism and measurement of the heat flows that can be explained by mantle convection.

<u>Prerequisite 11: The researcher has stubbornness in thinking and good networking with colleagues in the field</u>

As a further prerequisite for paradigm change in a field, we want to mention the willingness of researchers to go their own way (if need be, and not at all costs). An overly strong orientation towards established persons and expert opinions and a lack of willingness and courage to consider and examine alternatives can get in the way of change. To break up existing paradigms, what is needed are people who think out of the box, not opportunists or careerists. Good examples of courageous researchers are Hubble (1929) and Hubble and Humason (1931) in the area of cosmology, Vine and Matthews (1963) in geophysics, and Einstein (1905b, 1915) and Planck (1900a, 1900b) in quantum physics. In addition to a healthy portion of stubbornness in thinking, scientific research that leads to paradigm change is hardly conceivable without the researcher's integration in a research group as an institutional and intellectual home as well as in an extended network of colleagues in the field with whom information can be exchanged and problems and findings discussed. Significant prerequisites for very successful science are a group atmosphere that fosters scientific interests, a flat hierarchy makes communication possible across age and status differences, and constructive criticism and simulating encouragement (Kumar, 2008; Oreskes, 2003).

Summary

Kuhn (1977) named the fundamental prerequisites for the changing of an old theory to a new theory (see above) but did not describe what the necessary prerequisites of scientific revolution are. Paradigm change needed to be examined further – especially in view of the complexity of modern research. For this reason we propose the AKP as an addition to or refinement of Kuhn's (1962) theory as an invitation to identify the precise set of prerequisites (some possible prerequisites are mentioned above). In that the AKP can break down the process towards a revolution in greater detail, it does more justice to the complexity of the course of scientific progress today. The AKP does not deliver any new or previously overlooked conditions for scientific revolutions; instead, it provides a new perspective, an ordering schema, for revolutions. Prerequisites 1, 3, and 5 above for a scientific revolution correspond approximately to the first two of the five characteristics of a good scientific theory named by Kuhn (1977, p. 322): accuracy and consistency. Prerequisites 6, 7 and 8 above correspond to two more of the characteristics that Kuhn (1977) said a theory should have: simplicity and fruitfulness. Also prerequisite 8 above, explanatory power, was called the same thing and discussed by Kuhn (1977).

The AKP based on the prerequisites presented here and their step by step fulfillment should not be seen too statically. Depending on discipline and time period, the one or other prerequisite will carry more or less weight. For example, solid evidence played a decisive role in cosmology (example 1) and geophysics (example 2) and an important but not decisive role in the development of quantum physics (example 3). Techniques for verification were also of rather secondary importance for quantum physics, and up to today there is still no explanation of the quantization of physical properties. All in all, however, we can assume that most of the prerequisites are necessary conditions for a paradigm change but that no one prerequisite alone is sufficient. Only when all important prerequisites are fulfilled will there be a scientific breakthrough.

In a case study taken from a modern area of research, Holton, Chang, and Jurkowitz (1996) discussed a schema similar to the AKP with concrete prerequisites like the ones above and:

> set forth an account of the discovery of the first high-temperature superconductors as Müller and Bednorz experienced it, with particular attention to the resources, either intellectual or material, on which the discovery depended. Then, based on this narrative … [they] put forward a systematic analysis, a schema designed to help answer the more general question of what it takes to make a scientific advance. (p. 366)

However, the research area of high temperature superconductivity is not yet completed, insofar as no generally recognized mechanism of the phenomenon has been found and the discovery of superconductors must be seen as more of an intermediary step. For this reason, a schema developed based on this example cannot be complete and generalizable. But there are certainly similarities to the AKP as it is proposed in this paper.

## 5  Discussion

In this paper we presented the AKP as a concept that can used to explain success in science: Success in science always depends on several key aspects that must all be fulfilled. One missing aspect out of many other aspects will lead to failure and makes the failure a unique matter. With this definition of success and failure, the AKP follows an integrative approach that assumes that phenomena (here, success in science) are complex and multidimensional. For instance, with the AKP the two most important citation theories can be put in a broader context (see above). With the integrative approach the AKP contradicts common notions of success, in that success is not attributed to special characteristics of the successful person or thing. Following the AKP, what is unique is the unsuccessful thing, for it is unsuccessful in a very particular way. With this simple concept, the AKP does not introduce previously overlooked, fully new conditions for scientific success but instead provides an ordering principle for the (many) conditions for success.

The implicit demand of the AKP is to seek several conditions and not one condition for success in science. This is helpful not only for description and explanation of a particular success (such as achieving a scientific breakthrough in a field); it also encourages us to think about the general conditions for success in many areas of scientific work (such as the assessment of manuscripts submitted to journals). For instance, what criteria must an application for a scholarship or grant fulfill to be successful? At many agencies awarding scholarships, these criteria are not stated explicitly and transparently – that is, they are not known to the reviewers engaged by the organization to review the applications. Only if the conditions under which an application to a funding organization can be successful are stated transparently can the validity of the selection decisions be checked afterwards (see here Popper, 1962). That is, only then can the extent to which the AKP is effective as a concept be checked.

A suitable technique to check the effectiveness of the AKP in a particular empirical situation is the Boolean probit statistical technique (Braumoeller, 2003). Using this technique it can be examined whether all factors held to be the cause of an event are given with high probability in an empirical situation for the event to occur. As we mentioned above, Bornmann and Daniel (2005) used this technique to examine the selection process at an international foundation awarding research fellowships. For that selection process it was found that the AKP was indeed in effect, as for the success of an application (success being the award of a fellowship) all predetermined criteria (positive assessment of the research project, in conjunction with both a positive assessment of the laboratory in question and a positive assessment of the applicant's achievements by reviewers) had to be fulfilled. If at least one assessment of an application is negative, the application for a fellowship will not be approved and is not successful in a particular way. Following the study by Bornmann and Daniel (2005), the Boolean probit statistical technique could also be used in areas other than peer review – such as citations or scientific progress – to test whether the AKP is effective, that is, whether all factors defined in the run-up to something were in place for an event to occur (such as especially high citation rates for research papers or a scientific revolution).

If the Boolean probit technique is used to examine the AKP in the area of peer review, it should also be checked whether a selection of submissions is made according to the configuration model, meaning that only those manuscripts are selected that achieve a minimum rating on all criteria. A selection process that is not probable in science due to the generally scarce resources but is still a possible alternative would be to use the compensation model, where a negative rating on one criterion could be made up for by a positive rating on another criterion. If a process uses a mixed model, certain minimum ratings on the individual criteria must be achieved before any one criterion can be seen as compensation for another.

In this paper, in addition to peer review we took a look at citation counts and scientific progress – that is, at areas in science that are closely connected:

> New knowledge claims – the results of scientific practices and research activities – are
> formulated in texts that are validated through the review process and eventually
> accepted for publication (or not). If published the new knowledge claims can be
> integrated into bodies of scientific knowledge and eventually become part of a

> structuring global repertoire of scientific knowledge. (Lucio-Arias & Leydesdorff, 2009, p. 265)

For all three of those areas, we named factors, or prerequisites, that should be in place for a successful event to occur. Even though based on previously published literature on research in these areas we tried to name important key aspects, we are conscious of the fact that with great probability not all of the aspects that we outlined here are in fact key aspects and further key aspects may exist. But it was not our intention in this paper to draw up an exhaustive list of these key aspects but instead to introduce the AKP as a new concept in science studies and to explain its essential features. The task in future research in these three areas should be to identify the prevailing important conditions for success. Also not in connection with the AKP, in all three areas there is great interest in identifying these factors.